\documentclass{INTERSPEECH2023}

\interspeechcameraready

\title{Scaling Laws for Discriminative Speech Recognition Rescoring Models}
\name{Yile Gu, Prashanth Gurunath Shivakumar, Jari Kolehmainen, Ankur Gandhe, Ariya Rastrow, and Ivan Bulyko}
\address{
  Amazon Alexa}
\email{ \{yilegu,psshvak,jkolehm,aggandhe,arastrow,ibbulyko\}@amazon.com}

\begin{document}

\maketitle
 
\begin{abstract}
Recent studies have found that model performance has a smooth power-law relationship, or scaling laws, with training data and model size, for a wide range of problems. These scaling laws allow one to choose nearly optimal data and model sizes. We study whether this scaling property is also applicable to second-pass rescoring, which is an important component of speech recognition systems. We focus on RescoreBERT as the rescoring model, which uses a pre-trained Transformer-based architecture fined tuned with an ASR discriminative loss. Using such a rescoring model, we show that the word error rate (WER) follows a scaling law for over two orders of magnitude as training data and model size increase. In addition, it is found that a pre-trained model would require less data than a randomly initialized model of the same size, representing effective data transferred from pre-training step. This effective data transferred is found to also follow a scaling law with the data and model size.

\end{abstract}

\section{Introduction}

State-of-the-art automatic speech recognition systems (ASR) perform second-pass rescoring in which the n-best hypotheses, generated by the first-pass, are reranked to improve accuracy~\cite{deliberation-nips,two-pass,deliberation2020,hu2021transformer,ankur}. The need for second-pass rescoring stems from the architectural constraints of running a low-latency and streaming first-pass. To ensure improved performance and better WER, minimum WER (MWER) loss is typically applied~\cite{mwer2016,deliberation2020,ankur,mwer2018} when training a second-pass rescoring model.

Research has found that model performance has predictable and favorable scaling properties with respect to training data size, model size, and compute across a variety of modalities, including language~\cite{kaplan2020scaling, ghorbani2021scaling,gordon2021data}, vision~\cite{henighan2020scaling, zhai2022scaling}, and acoustics~\cite{droppo2021scaling}. These scaling laws not only provide supportive evidence in favor of “large models”, but they also provide researchers with the ability to determine the suitable configuration based on training data, model size, and computation.

In spite of some recent studies showing that increasing model size results in better ASR rescoring performance~\cite{oraclesearch,hu2023scaling,xu2022rescorebert,huang2019empirical}, there has not yet been a systematic study on the scaling properties of second-pass rescoring with data and model size. Understanding the scaling properties of adopting optimal data and model size would be critical, as (a) acquiring the transcribed data required for discriminative training is costly, and (b) when deployed in production, larger models require more expensive hardware to maintain low latency.

Additionally, pre-training’s impact on ASR rescoring performance is not well understood. In spite of their success at reducing the amount of annotated data needed for downstream language understanding tasks, pre-trained language models, such as BERT~\cite{bert}, have not been widely applied to second-pass ASR rescoring. Recent studies have examined how pre-trained models can improve ASR rescoring~\cite{mlm-scoring1,mlm-scoring2,wu22_interspeech, xu2022rescorebert,hu2022improving,futami2021asr,fohr2021bert}, but there is no systematic study of how pre-training affects rescoring performance for different models and data sizes, and whether a scaling law, proposed in a previous work ~\cite{hernandez2021scaling}, can capture the effect.

In this work, we fill the gaps in the literature regarding rescoring models employed in speech recognition systems, and systematically study the scaling properties for both randomly initialized and pre-trained discriminative rescoring models. We use the recently proposed RescoreBERT~\cite{xu2022rescorebert} rescoring model, which is based on a BERT architecture pre-trained on large corpora. The RescoreBERT model encodes the full context of the hypothesis using a bi-directional self-attention architecture. We demonstrate that WER follows a power-law relationship with training data size and model size for over two orders of magnitude of the range studied. Furthermore, we underscore the importance of pre-training for second-pass rescoring, and show that effective data transferred from pre-training allows a model to require less training data to achieve the same performance. A power-law relationship can also be used to describe the effective data transferred.

\section{Experimental Setup}

\subsection{RescoreBERT}
\begin{figure}[t]
\centering
\includegraphics[width=7cm]{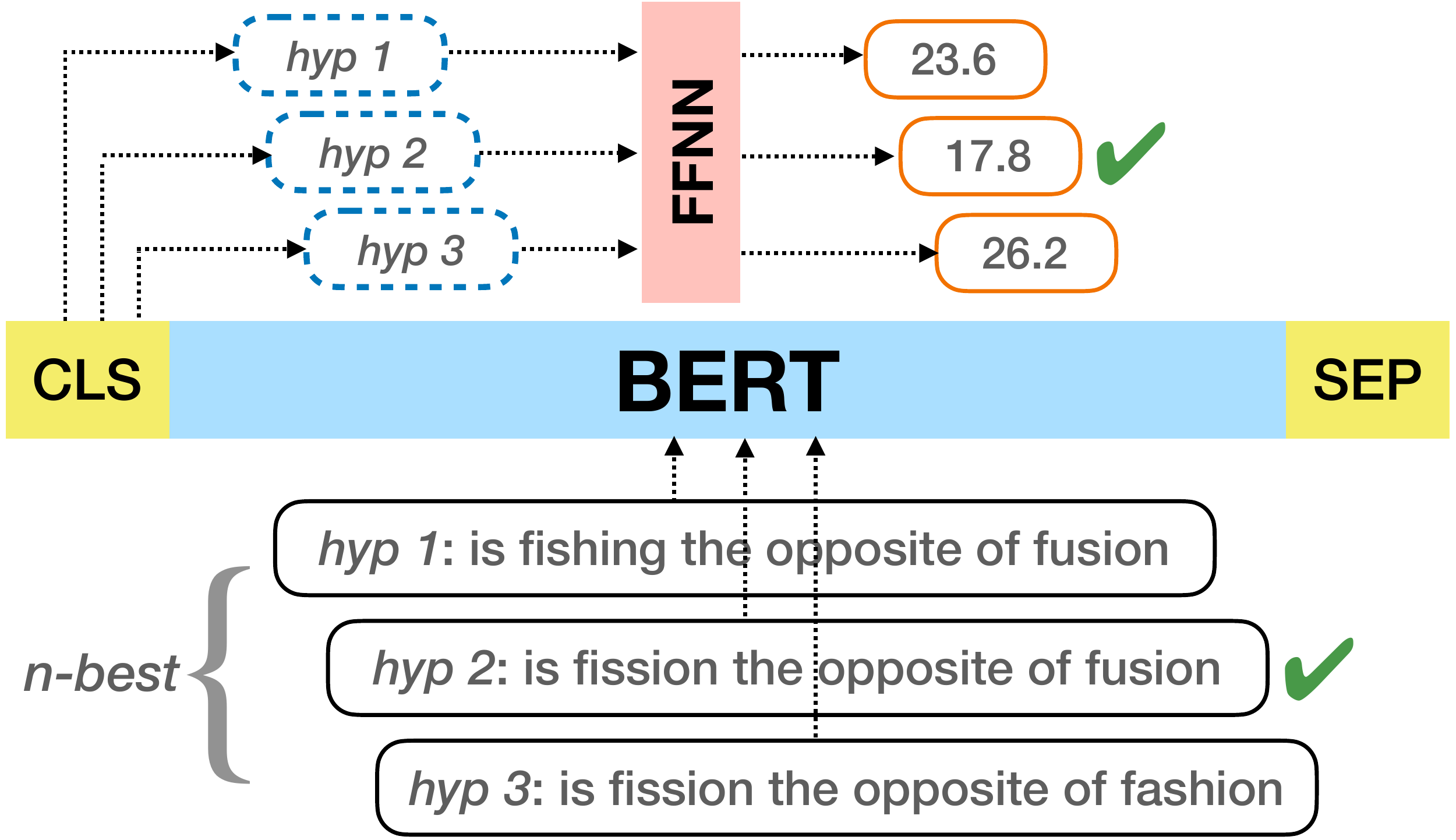}
\caption{Illustration of RescoreBERT. Each hypothesis is individually encoded by BERT and represented by CLS; it is then followed by a feed-forward neural network to compute a sentence-level second-pass LM score. The scores are then interpolated with first pass scores for re-ranking. The figure is reproduced from the original RescoreBERT paper~\cite{xu2022rescorebert}.}
\label{fig:rescorebert}
\vspace{-2ex}
\end{figure}

We use RescoreBERT~\cite{xu2022rescorebert} as the rescoring model to explore the scaling law. As illustrated in Figure~\ref{fig:rescorebert}, RescoreBERT model uses a BERT model with a feed-forward network attached at the BERT encoder classification (CLS) token embedding (CLS) to predict a single second-pass score for a given n-best hypothesis. This score is linearly interpolated with first pass score for re-ranking. The final score $s_i$ used for re-ranking is therefore,
\begin{align}
s_i=s_i^f+w\cdot s_i^s, 
\end{align}
where $s_i^f$ and $s_i^s$ are the scores from first and second passes, respectively, and w is a hyper-parameter. 

Following~\cite{xu2022rescorebert}, we train a RescoreBERT model using a minimum WER (MWER) discriminative loss~\cite{mwer2018}. The training minimizes expected word error rate calculated on given n-best hypotheses:
\begin{align}
    P_i &= \frac{e^{-s_i}}{\sum_{j=1}^n e^{-s_j}} \\
    \overline{\epsilon}_H &= \frac{1}{n} \sum_{i=1}^n \epsilon_i \\
    \mathcal{L}_{\text{MWER}} &= \sum_{i=1}^n P_i \cdot (\epsilon_i - \overline{\epsilon}_H) \label{eq:mwer-loss}.
\end{align}
$P_i$ is the posterior probability of a hypothesis $i$, and $\epsilon_i$ is the edit distance from the ground truth transcription. The MWER loss $\mathcal{L}_{\text{MWER}}$ represents the expected number of relative word errors, with $\overline{\epsilon}_H$ being the averaged word errors across the n-best list.

\subsection{Experiments Performed}
In the original RescoreBERT paper~\cite{xu2022rescorebert}, discriminative training is performed as a fine-tuning step using pre-trained BERT models. In this study, we train both pre-trained and randomly initialized models to study the effects of pre-training. 

The effective batch sizes used for training are 512, 512, 512, 2048 sets of n-best hypotheses for the 5M/17M/170M/700M models, respectively. Learning-rate decay and Adam optimizer with default parameters are used, with initial learning rates of $10^{-5}$ for 5M/17M/170M and $5\times10^{-6}$ for 700M models. Training is applied with different model sizes and data sizes. We apply early stopping based on a development set, and report WER based on the test set.


\subsection{Model and Data Sizes}
We use four variants of BERT models with different sizes as outlined in Table~\ref{tab:models}. The models are pre-trained with MLM (Masked Language Model) objective, first with Wikipedia and mC4~\cite{soltan2022alexatm}, and then internal catalog data (up to around 1 trillion tokens in total pre-training data).

\begin{table}[h]
    \centering
    \caption{Summary of architecture details of BERT models.}
    \label{tab:models}
    \begin{tabular}{m{30mm}|c|c|c|c}
    \hline
        Model parameters (excluding embeddings) & $5$M & $17$M & $170$M & $700$M \\
        \hline
        Hidden Size & $320$ & $768$ & $1024$ & $1536$\\
        Number of Layers & $4$ & $4$ & $16$ &$20$ \\
        Number of Attention Heads &$16$& $16$ &$16$ &$16$ \\
        Intermediate Layer Dimension &$1200$ & $1200$ & $3072$ & $6144$ \\
        Embedding Parameters & $ 49$M & $118$M & $157$M & $460$M \\
        \hline
    \end{tabular}
\end{table}

In the discriminative training phase of RescoreBERT models, we utilize internal datasets consisting of de-identified user interactions with a conversational agent in English. We use an RNN-T model~\cite{he2019streaming} as the first-pass model to generate n-best hypotheses. The train, dev and test splits, in this study, contain 95300, 30 and 10  of utterances, respectively.
To study the effects of training data, different fractions of training are used. We use internal data for this study, as there is no readily available public speech data at this large size to study scaling laws. However, the authors believe that the phenomena described in this paper should apply to any similar set of speech data. 

\section{Results and Discussion}

\subsection{Normalized WER (Word Error Rate)}
To showcase the effectiveness of second-pass rescoring, we report the results in $\mathrm{WER_{norm}}$, which is defined as
\begin{align}
    \mathrm{WER_{norm}} = \frac{\mathrm{WER_{2P}}-\mathrm{WER_{oracle}}}
    {\mathrm{WER_{1P}}-\mathrm{WER_{oracle}}} \label{eqn:norm}
\end{align}
where $\mathrm{WER_{1P}}$ and $\mathrm{WER_{2P}}$ are WER before and after second-pass rescoring. $\mathrm{WER_{oracle}}$ is oracle WER, which provides a lower bound for the minimum WER achievable from second-pass rescoring. For the test in this study, $\mathrm{WER_{oracle}}$ is $66\%$ relative lower than $\mathrm{WER_{1P}}$, meaning that if $\mathrm{WER_{norm}}$ is 0.5, second-pass rescoring would provide $33\%$ relative reduction in WER over first pass. $\mathrm{WER_{norm}}$ becomes $0$ when WER from rescoring approaches oracle, and becomes $1.0$ when rescoring fails to improve over first pass.

\begin{figure}
  \centering
         \includegraphics[width=1.1\linewidth]{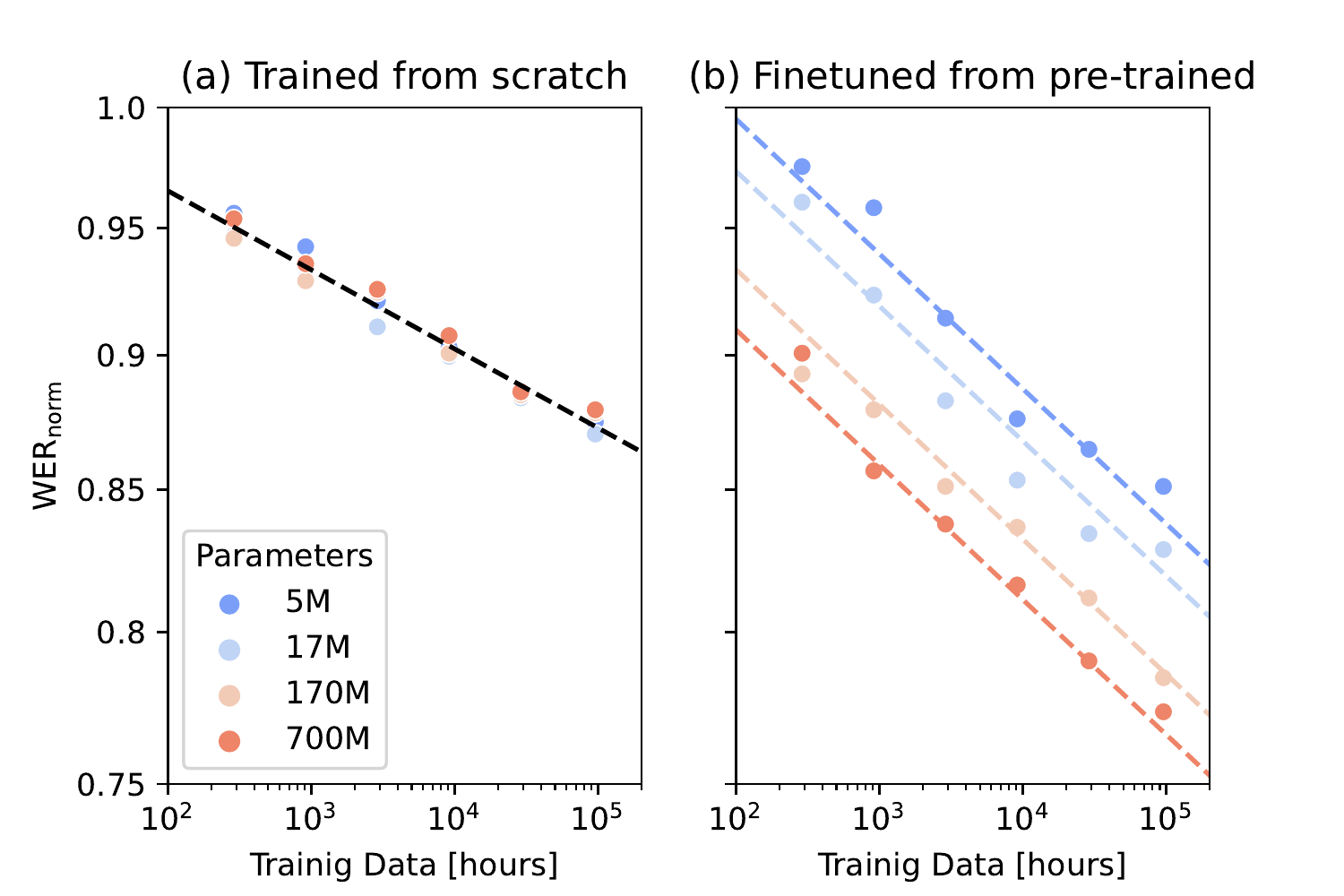}
    \caption{Normalized WER $\mathrm{WER_{norm}}$ (defined in Equation~\eqref{eqn:norm}) versus training data for (a) models trained from scratch (b) models finetuned from pre-trained models.Dashed lines denote model predictions from Equation~\eqref{eqn:non} for (a) and Equation~\eqref{eqn:pre} for (b). Both axes are on log scale.}
    \label{fig:training_data}
\end{figure}

\begin{figure}
  \centering
         \includegraphics[width=1.1\linewidth]{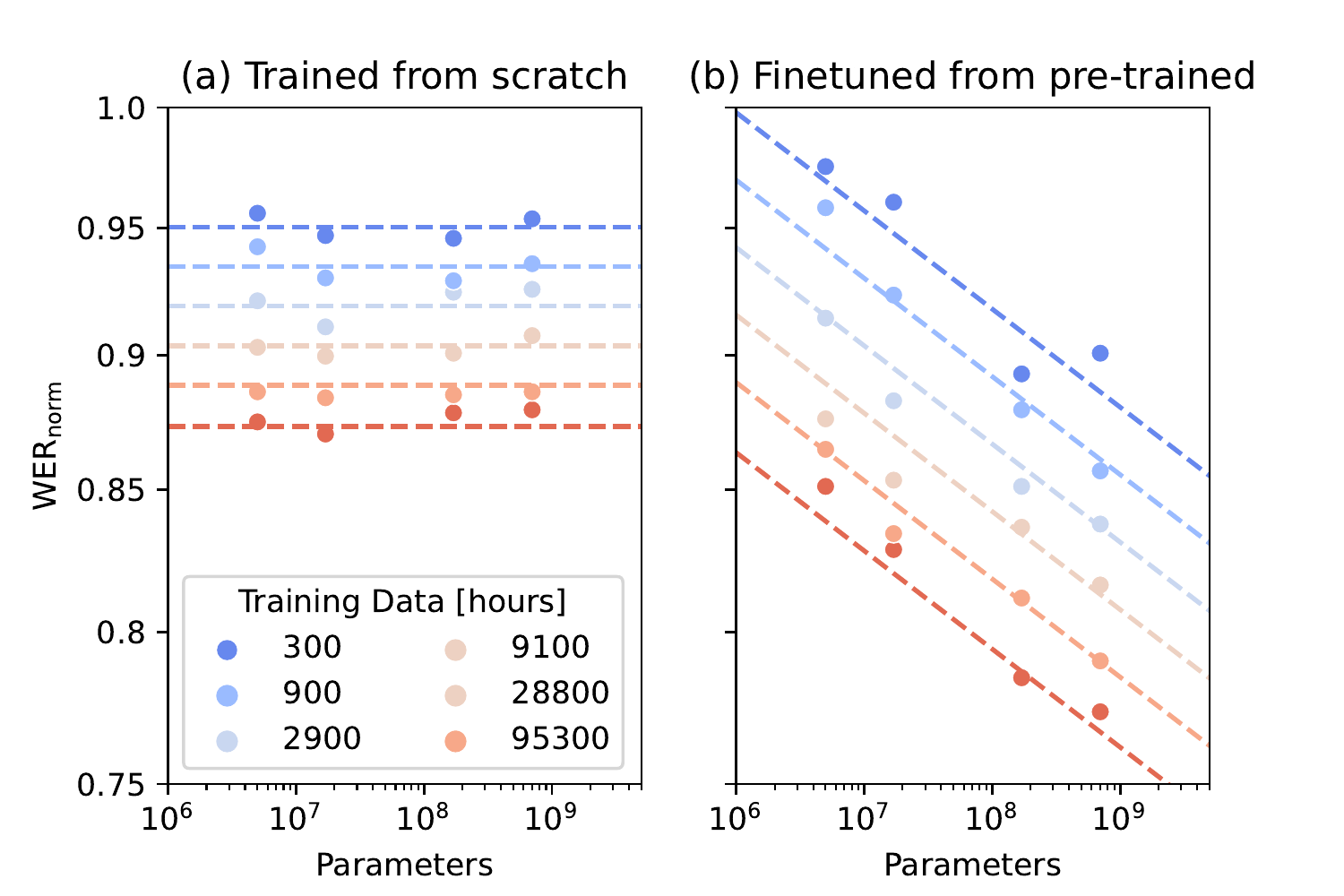}
    \caption{Same data as in Figure~\ref{fig:training_data} but plotted with a different x-axis.  Normalized WER $\mathrm{WER_{norm}}$ (defined in Equation~\eqref{eqn:norm}) versus model parameters for (a) models trained from scratch (b) models finetuned from pre-trained models. Dashed lines denote model predictions from Equation~\eqref{eqn:non} for (a) and Equation~\eqref{eqn:pre} for (b). Both axes are on log scale.}
    \label{fig:parameters}
\end{figure}

\subsection{Training from Scratch}\label{section_scratch}
We first study the effects of training data and model sizes on WER when trained from scratch. The results can be found in Figures~\ref{fig:training_data}(a) and~\ref{fig:parameters}(a). We observe that $\mathrm{WER_{norm}}$ is independent from model sizes, and has a power-law relationship with training data. It indicates that for the range of data in this study, the model is limited by training data. Because the training data here require human annotation and are hence limited, this study is limited to this low-data regime, but we expect that this regime would be the case for most speech recognition applications; while a model size of $5$M is at the lower end of what's typical for a BERT model, $95,300$  of training data are at the upper end of what's usually used for speech recognition systems. It is in contrast to previous work on language modeling~\cite{kaplan2020scaling} where text data is easily accessible, which allows it to explore both scenarios of being limited by training data and limited by model sizes. 

For this regime where data is limited, the previous work~\cite{kaplan2020scaling} proposed the following equation to describe the relationship between test set performance and training data size, 
\begin{align}
    L=\left(\frac{D_C}{D}\right)^{\alpha_D} \label{eqn:literature_non},
\end{align}
where $L$ is the test loss, $D$ is training data size, and $D_C$ is a model parameter and represents the critical value of $D$ where the contribution of data to the loss function is equal to $1.0$.

We found that this equation also describes the data well for $\mathrm{WER_{norm}}$, as evidenced by the goodness of the fit in dashed lines in Figures~\ref{fig:training_data}(a) and~\ref{fig:parameters}(a). The equation is now,
\begin{align}
    \mathrm{WER_{norm}}=\left(\frac{D_C}{D}\right)^{\alpha_D} \label{eqn:non},
\end{align}
where $D_C$ is 8.82 and $\alpha_D$ is 0.0146, both of which are estimated from fitting the data of $log(WER_{norm})$ vs $log(D_C)$. 
The small value for $D_C$ indicates that a small amount of data is sufficient to bring $\mathrm{WER_{norm}}$ below 1.0. This makes sense for our data, since if the second pass just outputs the same scores for all the n-best hypotheses, due to linear interpolation with the first pass scores, WER would be the same as the first pass, and $\mathrm{WER_{norm}}$ would be equal to $1.0$.

\subsection{Finetuned from Pre-trained Models}
We then study how WER is changed when the model is finetuned from a pre-trained model. As shown in Figures~\ref{fig:training_data} and~\ref{fig:parameters} pre-training model helps reduce WER as expected.

As shown in Figure~\ref{fig:parameters}(b), a larger model would now lead to more improvement, as opposed to what is found for models trained from scratch where the model performance is independent to the model size (as in Figure~\ref{fig:parameters}(a)). It shows that for pre-trained models, the improvement from increasing the model size is not due to the larger capacity of the model to better learn from finetuning data, but rather it is its ability to better memorize and leverage the knowledge from pre-training. Without studying the scaling laws of both randomly initialized and pre-trained models, it would be hard to distinguish the effects from the two, underscoring the importance of such exercises.

In Figure~\ref{fig:training_data}, to achieve the same WER for a given model size, one would require less training data due to pre-training. For example, for a 5M model to achieve $\mathrm{WER_{norm}}$ of 0.875, data required would be $\sim 30$k  if trained from scratch, but only $\sim 1$k hour if pre-trained. Furthermore, this delta in data required depends on model size (as discussed in the previous paragraph) and also training data itself (reflected by the change of slope between Figures~\ref{fig:training_data} (a) and (b)). 

This reduction in data required due to pre-training and its relationship to finetuning data and model size are consistent with a previous work~\cite{hernandez2021scaling} studying scaling for transfer learning. In the paper, it introduces the concept of effective data transferred from pre-training, $D_T$, to capture this delta, and found that it can be captured by $D_T=k D^\alpha N^\beta$. Hence, for a given data size, due to pre-training, its effective data size would become $(D+D_T)$, or $(D+k D^\alpha N^\beta)$. 

As also explained in the paper~\cite{hernandez2021scaling}, in this low data regime, one can ignore the contribution of the original $D$ as the effective data from transfer is much greater than the amount of data finetuned on, $D_T\gg D$, which is also demonstrated in the example before where $D_T$ and $D$ are $30$k and $1$k  respectively.

Hence, the effective data size $(D+k D^\alpha N^\beta)$ can be simplified to $k D^\alpha N^\beta$. Plugging into Equation~\eqref{eqn:non}, and we have the equation as follows,
\begin{align}
    \mathrm{WER_{norm}}=\left(\frac{D_C}{k D^\alpha N^\beta}\right)^{\alpha_D} \label{eqn:pre}.
\end{align}
$D_C$ and $\alpha_D$ have the same values as in Equation~\eqref{eqn:non}.
As shown with dashed lines in Figures~\ref{fig:training_data}(b) and~\ref{fig:parameters}(b), the equation captures the data well, where $k$ is $2.27\times10^{-11}$, $\alpha$ is 1.71, and $\beta$ is 1.24. In contrast to the previous work~\cite{hernandez2021scaling} where the pre-training and finetuning steps are trained with the same loss and different domains of data, it shows that this equation also holds even when the pre-training and finetuning have different training objectives. The ratio of $\alpha$ and $\beta$ suggest that a 10x increase in model size would be worth approximately a 5.3x increase in data size. Hence, the scaling law here offers useful insights as one decides between obtaining more training data and increasing model sizes to improve the model performance.




\subsection{Applicability of the Scaling Law}
Hence, we have proposed scaling laws for both randomly initialized model and pre-trained models as in Equations~\eqref{eqn:non} and \eqref{eqn:pre}, respectively. Even though the data sizes (from $300$  to $95300$ ) and model sizes (from $5$M to $700$M) represent most speech recognition systems, it is still helpful to discuss the limitation in the applicability of the scaling laws described here. 

First, as also described in the previous paper~\cite{hestness2017deep}, scaling laws would break down at both extreme ends of the data size spectrum. When there is very little data, the model can perform as well as random guessing, and in our case the second pass cannot further improve over the first pass, yielding 1.0 for $\mathrm{WER_{norm}}$; at the other extreme, there would be a non-zero lower bound error past which the models will not be able to improve with more data (or model parameters). This lower bound error or irreducible error would include Bayers error and noises in the data.

Second, as shown in Section~\ref{section_scratch}, for randomly initialized models, the regime in this study is limited by the amount of data. Once we increase by a sufficient amount of data, the model size would start to influence $\mathrm{WER_{norm}}$ even for randomly initialized models, and scaling laws for both randomly initialized models and pre-trained models likely need to be modified, as shown in the previous paper~\cite{hernandez2021scaling}.

\section{Conclusions}
Using RescoreBERT as the rescorer, we demonstrate that scaling laws are also applicable for discriminative speech recognition rescoring models, for over two orders of magnitude of range studied. For randomly initialized models, WER is found to have power-law relationship with training data size, and independent from model sizes, indicating that it is operating at a regime where the data is limited. Due to the large data in this study (almost $100$k ), we expect that this regime would be the case for most speech recognition applications.

For pre-trained models, it is found that WER also has a power-law relationship with training data and model sizes. Noticeably, different from randomly initialized models, WER now decreases with increasing model size, as larger model has the ability to better memorize and leverage the knowledge from pre-training. In addition, it is found that a pre-trained model would require less data than a randomly initialized model of the same size, representing the effective data transferred from pre-training. This effective data transferred also follows a scaling law with the data and model sizes.

While we focus on RescoreBERT as the discrminative rescoring model to study the scaling laws, it would be interesting to see whether the scaling laws proposed in this study hold true for other types of discriminative rescoring models, especially when they are conditioned on first-pass audio/lattice~\cite{deliberation2020,hu2021transformer,ankur,pandey2022lattention}.


\newpage
\bibliographystyle{IEEEtran}
\bibliography{mybib}

\end{document}